%% file: DASH-SIM.tex
 \documentclass[sigconf]{acmart}

\usepackage{booktabs} 
\usepackage{xcolor}

\usepackage{amsmath}

\usepackage[ruled, linesnumbered]{algorithm2e}
\usepackage{adjustbox}
\usepackage{graphicx}
\usepackage{subcaption}
\usepackage{soul,xcolor}
\usepackage{graphicx}
\usepackage{multirow}
\usepackage{rotating}
\usepackage{array}
\usepackage{enumitem}
\usepackage[normalem]{ulem}
\usepackage{dblfloatfix}
\usepackage[skip=3pt]{caption}

\definecolor{OliveGreen}{rgb}{0,0.6,0}
\definecolor{skyblue}{rgb}{0.06, 0.75, 0.99}

\setstcolor{red}
\setcopyright{acmcopyright}

\copyrightyear{2019}
\acmYear{2019}
\acmConference[CODES/ISSS '19]{International Conference on Hardware/Software Codesign
and System Synthesis}{October 13--18, 2019}{New York, NY, USA}
\acmBooktitle{International Conference on Hardware/Software Codesign and System Synthesis
(CODES/ISSS '19), October 13--18, 2019, New York, NY, USA}
\acmPrice{15.00}
\acmDOI{10.1145/3349567.3351719}
\acmISBN{978-1-4503-6923-7/19/10}

\usepackage{multirow}
\usepackage[acronym]{glossaries}
\newacronym{tx}{WiFi-TX}{WiFi transmitter}

\begin{document}
	
	\title{\vspace{-2mm} Work-in-Progress: A Simulation Framework for Domain-Specific System-on-Chips\vspace{-3mm}}
	\author{
    Samet E. Arda$^1$, Anish NK$^1$, A. Alper Goksoy$^1$, Joshua Mack$^2$, Nirmal Kumbhare$^2$, \\ Anderson L. Sartor$^3$, Ali Akoglu$^2$, Radu Marculescu$^3$ and Umit Y. Ogras$^1$
    \\
    \large{$^1$School of Electrical Computer and Energy Engineering, Arizona State
    University, Tempe, AZ, USA \\ $^2$Electrical and Computer Engineering, The University of Arizona, Tucson, AZ, USA \\ $^3$Electrical and Computer Engineering, Carnegie Mellon University, Pittsburgh, PA, USA}
    }

 	\input{files/0-abstract.tex}
    \maketitle

	\input{files/1-introduction.tex}

	\input{files/2-overview.tex}
	\input{files/3-case_studies.tex}

	\bibliographystyle{abbrv}
	\bibliography{references/references}
	
\end{document}

%% file: files/0-abstract.tex




%% file: files/1-introduction.tex
 \vspace{-4mm}
 \section{Introduction and Background}

Homogeneous general purpose processors provide flexibility to implement a variety of applications and facilitate programmability. 
In contrast, heterogeneous system-on-chips (SoCs) that combine general purpose and specialized processors offer great potential to achieve higher efficiency while maintaining programming flexibility. In particular, domain-specific SoCs (DSSoC), a class of heterogeneous architectures, tailor the 
architecture and processing elements (PE) to a specific domain. Hence, they can provide superior energy-efficiency compared to general purpose processors by exploiting the characteristics of target applications.


DSSoCs can fulfill their potential only if they integrate the set of accelerators required by the target domain and utilize these resources effectively. 
Therefore, the first step in the design of DSSoCs is analyzing the domain applications to identify the most commonly used kernels. 
This analysis is necessary to determine the set of hardware accelerators in the architecture.
For example, a DSSoC that targets wireless communications domain will most likely have Fourier transform (FFT) accelerators.
The next step is to design the DSSoC, including the PEs and network-on-chip that interconnects them.
Then, a wide range of design- and run-time algorithms are employed to schedule the domain applications to the PEs in the DSSoC~\cite{smit2005run, de2010dynamic}.
Similarly, dynamic voltage and frequency scaling (DVFS) policies, 
such as Ondemand and thermal management techniques have been applied to efficiently manage the power and temperature of SoCs~\cite{bhat2018algorithmic}.
However, existing approaches are typically evaluated in isolated  environments and different in-house tools.
Thus, there is a strong need for a unified simulation environment to enable design space exploration and dynamic resource management of domain applications.

StarPU framework provides the ability to perform run-time scheduling and execution management for directed acyclic graph (DAG) based programs on heterogeneous architectures~\cite{augonnet2011starpu}. It is integrated into SimGrid \cite{casanova2013simgrid} framework. 
However, SimGrid was developed in the context of providing fast simulation for distributed systems, and the authors acknowledge that it is not intended to scale down to simulating real-time multi-threaded systems or comparing kernel schedulers and policies \cite{simgrid-limits}. 
A recent work~\cite{xiao2019self-optimizing} targets domain-specific programmability of heterogeneous architectures through intelligent compile- and run-time mapping of tasks across different PEs. 
In this study, the authors employ three different simulators. 
The proposed framework would benefit this and other similar studies by providing a single integrated simulation framework.

In this work, we present 
an integrated \textit{open-source} simulation framework capable of evaluating \textit{both scheduling and dynamic thermal-power management algorithms}. 
It addresses rapid \textit{system-level power, performance, and temperature exploration} of DSSoCs. 
Besides facilitating the design of new scheduling and dynamic thermal-power management (DTPM) algorithms, 
the proposed framework also features built-in DVFS governors deployed on commercial SoCs and analytical power, performance, and temperature models. 
Finally, the framework includes five reference applications from wireless communication and radar processing domains.
These applications are profiled on commercial heterogeneous SoC platforms and provided as a benchmark suite integrated to our framework. 



%% file: files/2-overview.tex
\vspace{-3mm}
\section{Overview} \label{sec:goals}

\begin{figure}[t]
	\centering
 	\includegraphics[width=1\linewidth]{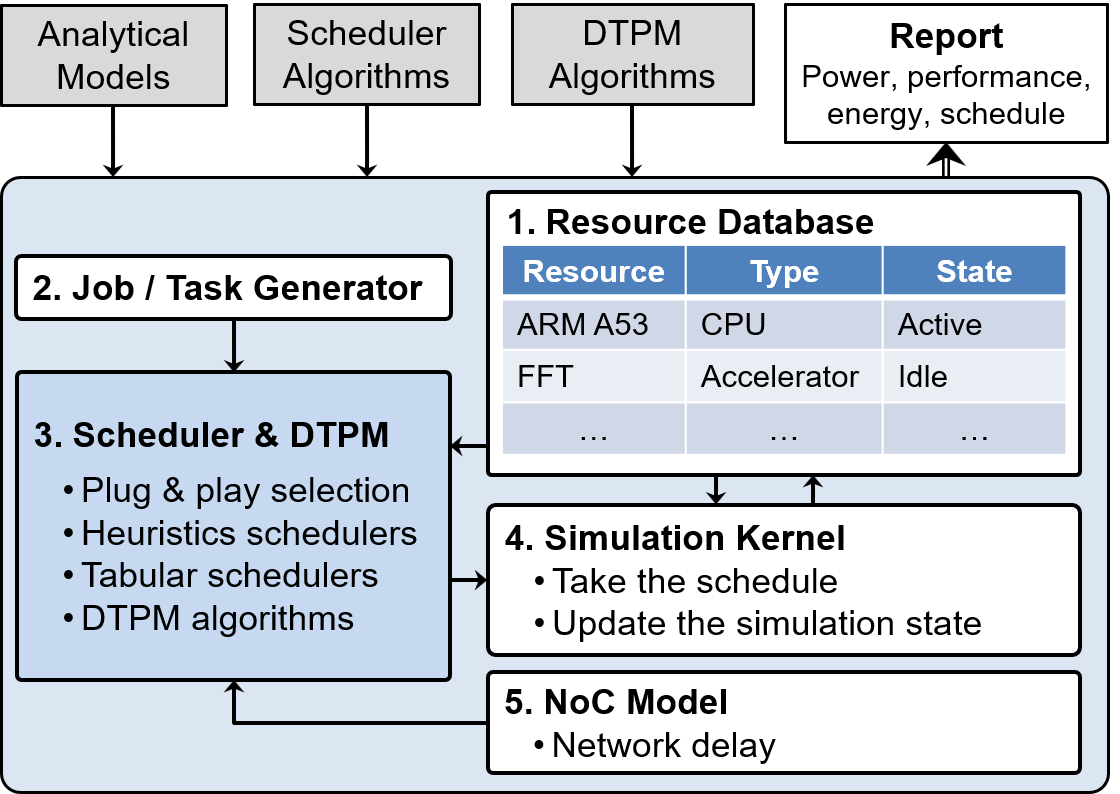}
	\caption{Simulation framework}
	\label{fig:DS3-framework}
\vspace{-4mm}
\end{figure}

The goal of the proposed simulation framework is to enable rapid development of scheduling and power management algorithms while enabling extensive DSSoC design space exploration. 

\begin{figure}[t]
\vspace{-2mm}
	\centering
	\includegraphics[width=0.9\linewidth]{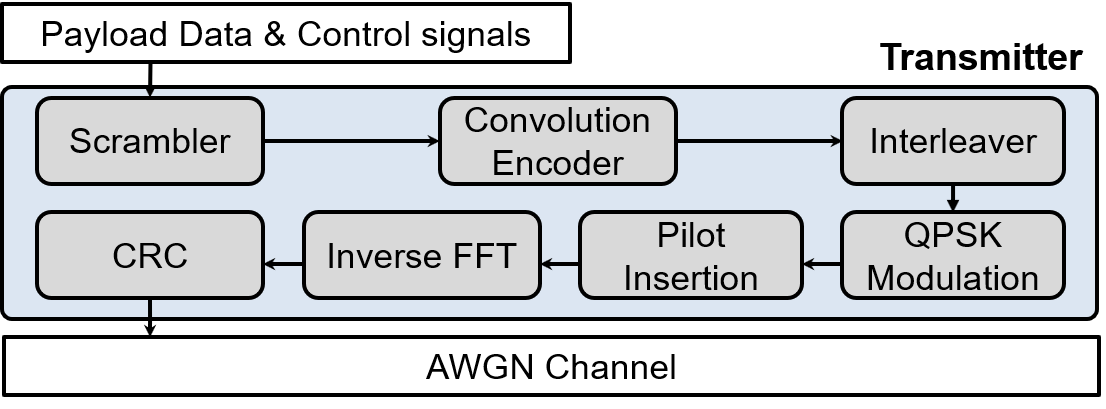}
	\caption{Block diagram of WiFi transmitter application}
	\label{fig:wifi_tx_bd}
\vspace{-4mm}
\end{figure}

The organization of the framework, designed to accomplish these objectives, is shown in Figure ~\ref{fig:DS3-framework}. 
The resource database contains the list of PEs along with expected latency of tasks in the application(s). 
The simulation is driven by the job generator
which injects instances of an applications to the simulator
following a given probability distribution. As an example, Figure ~\ref{fig:wifi_tx_bd} presents a block diagram for a WiFi transmitter (WiFi-TX) job that is composed of multiple tasks. The dependency among the tasks is represented using a DAG. 
Hence, the job generator produces the tasks for a transmitter job along with their dependencies. 

The simulation framework invokes the scheduler at every scheduling decision epoch with the list of tasks ready for execution.  
Then, the simulation kernel simulates task execution on the corresponding PE using execution time profiles obtained from our reference hardware implementations. 
Similarly, the framework employs analytical latency models to estimate interconnect delays on the SoC. 
After each scheduling decision, the simulation kernel updates the state of the simulation, which is used in subsequent decision epochs.

There are three built-in scheduling algorithms: 1) Minimum execution time (MET) scheduler~\cite{heuristic_comparison}, 2) Earliest task first (ETF) scheduler~\cite{workflow_schedule}, and 3) A table-based scheduler which can store any offline schedule, such as an assignment generated by an integer linear programming (ILP) solver, in the form of a look-up table. 
In addition, the framework enables a plug-and-play interface to choose between different scheduling algorithms. 
Hence, developers can implement their own algorithms and integrate them easily.

In parallel, power and energy estimates of each schedule are calculated by using power models~\cite{bhat2018algorithmic}. 
Using these models, the proposed framework aids the design space exploration of DTPM techniques. 
Similarly, the memory access and on-chip interconnect latency are modeled by the proposed framework. 
Finally, the framework generates plots and reports of schedule, performance, throughput, and energy consumption to aid users in analyzing the behaviour of various algorithms. 

%% file: files/3-case_studies.tex
\vspace{-4mm}
\section{Scheduling Case Study} \label{sec:case studies}


\renewcommand{\arraystretch}{0.9}
\begin{table}[b]
\vspace{-2mm}
\centering
\small
\caption{Execution profiles of WiFi-TX on Arm A7/A15 cores in Odroid-XU3 and hardware accelerators}
\label{tab:application_latencies}
\setlength\tabcolsep{3pt} 
\begin{tabular}{clclcc}
\hline
\multirow{2}{*}{\textbf{\begin{tabular}[c]{@{}c@{}}Sample \\ App.\end{tabular}}} & \multicolumn{1}{c}{\multirow{2}{*}{\textbf{Task}}} & \multicolumn{4}{c}{\textbf{Latency (\boldmath$\mu$s)}} \\ \cline{3-6} 
 & \multicolumn{1}{c}{} & \multicolumn{2}{c}{\textbf{\begin{tabular}[c]{@{}c@{}}HW Acc.\end{tabular}}} & \textbf{\begin{tabular}[c]{@{}c@{}}Odroid A7\end{tabular}} & \textbf{\begin{tabular}[c]{@{}c@{}}Odroid A15\end{tabular}} \\ \hline
\multirow{7}{*}{\textbf{WiFi-TX}} & 
\begin{tabular}[c]{@{}l@{}}Scrambler Enc.\end{tabular} & \multicolumn{2}{c}{8} & 22 & 10 \\
 \cline{2-6} 
 & Interleaver & \multicolumn{2}{c}{} & 10 & 4 \\ \cline{2-6} 
 & \begin{tabular}[c]{@{}l@{}}QPSK Modulation\end{tabular} & \multicolumn{2}{c}{} & 15 & 8 \\ \cline{2-6} 
 & Pilot Insertion & \multicolumn{2}{c}{} & 5 & 3 \\ \cline{2-6} 
 & Inverse-FFT & \multicolumn{2}{c}{16} & 296 & 118 \\ \cline{2-6} 
 & CRC & \multicolumn{2}{c}{} & 5 & 3 \\ \hline
\end{tabular}
\vspace{-4mm}
\end{table}

 \renewcommand{\arraystretch}{0.9}
 \begin{table}[b]
 \vspace{-1mm}
 \small
 \caption{SoC configuration for scheduling case studies}
 \vspace{-1mm}
 \label{tab:soc_configuration}
 \begin{tabular}{@{}llc@{}}
 \toprule
 \multicolumn{1}{c}{\textbf{Resource}} & \multicolumn{1}{c}{\textbf{Type}} & \multicolumn{1}{c}{\textbf{\# of Instances}} \\ \midrule
 Cortex-A15 & ARM big Architecture & 4 \\ 
 Cortex-A7 & ARM LITTLE Architecture & 4 \\
 \begin{tabular}[c]{@{}l@{}}Scrambler-Encoder\end{tabular} & Hardware Accelerator & 2 \\
 FFT & Hardware Accelerator & 4 \\
 \bottomrule
 \end{tabular}
 \end{table}

The proposed simulation framework enables evaluation of various scheduling algorithms for real-world applications with different DSSoC configurations. 
To this end, we developed reference designs for WiFi transmitter and receiver (WiFi-RX), low-power single-carrier, range detection, and pulse Doppler applications on two popular commercial heterogeneous SoC platforms, i.e., Xilinx Zynq ZCU-102~UltraScale MPSoC
and Odroid-XU3.
We profiled the task execution times for each application executed on these platforms. As an example, the execution times for different tasks in WiFi-TX are shown in Table~\ref{tab:application_latencies}. The SoC configuration chosen for the scheduling case studies are shown in Table~\ref{tab:soc_configuration}.

\begin{figure}[t]
\vspace{-4mm}
	\centering
	\includegraphics[width=1.0\linewidth]{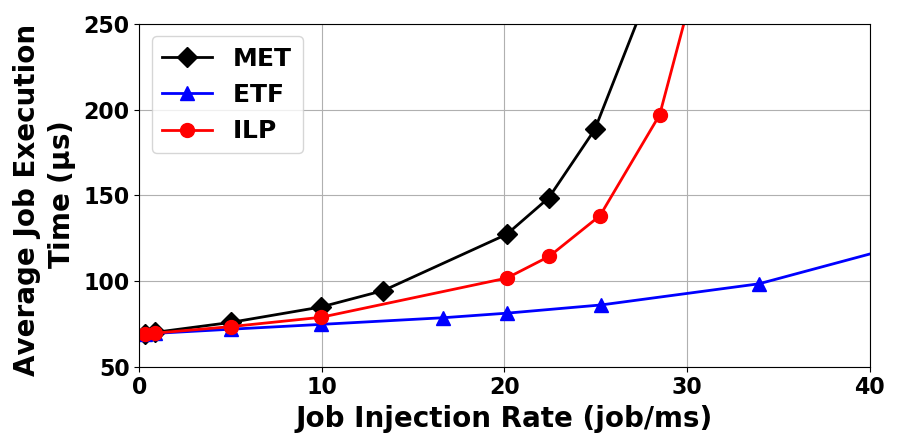}
	\vspace{-2mm}
	\caption{Results from different schedulers with a workload consisting of WiFi-TX jobs}
	\label{fig:tx_result}
\vspace{-3mm}
\end{figure}

In this study, the simulations run on an SoC configuration that mimics a typical heterogeneous SoC with a total of 14 general purpose cores and hardware accelerators. 
We schedule and execute the WiFi-TX task flow graph using the simulation framework and plot the average job execution time trend with respect to the job injection rate, as shown in Figure~\ref{fig:tx_result}.
To understand the performance of scheduling algorithms, we analyze the average execution time at varying rates of job injection.

All schedulers perform similar at low job injection rates (less than 5 job/$ms$). However, as the job injection rate\st{s} increases, the schedule from MET results in higher execution time since MET uses a naive representation of the system state for scheduling decisions, by only considering PEs with best execution times.
On the other hand, ILP uses a static table based schedule which is optimal for one job instance. 
At low injection rates (less than 5 job/$ms$), ILP provides a comparable schedule as jobs do not interleave.
However, as the injection rate increases, the ILP schedule is not optimal. 
The performance of ETF is superior in comparison to the others (see Figure~\ref{fig:tx_result})
since ETF utilizes the information about the communication cost between tasks and the current status of all PEs while making the scheduling decision.

To validate the proposed framework, we also implemented a subset of the scheduling algorithms on the Xilinx Zynq FPGA 
and then, compared the results for the applications in the benchmark suite with hardware measurements.
In summary, the experiment presented in Figure~\ref{fig:tx_result} demonstrates one of the many capabilities of the simulation environment. It allows the end user to evaluate workload scenarios exhaustively by sweeping the configuration space to determine the most suitable scheduling algorithm for a given SoC architecture.

\boldmath
\unboldmath
